# Warm Compressor systemOverview and status of the PIP-II cryogenic system

A Martinez[1], J Creus Prats[1], W Soyars[1], R Dhuley[1], B Hansen[1], Y Jia[1], A Chakravarty[2], M Goyal[2], T Banaszkiewicz[3], P Duda[3], M Stanclik[3]

[1]Fermi National Accelerator Laboratory
Batavia, IL 60510 USA

[2]Bhabha Atomic Research Center
Trombay, Mumbai 400084 India

[3]Wroclaw University of Science and Technology
Wroclaw, Poland

E-mail: martinez@fnal.gov

**Abstract.** The Proton Improvement Plan-II (PIP-II) is a major upgrade to the Fermilab accelerator complex, featuring a new 800-MeV Superconducting Radio-Frequency (SRF) linear accelerator (Linac) powering the accelerator complex to provide the world's most intense high-energy neutrino beam. The PIP-II Linac consists of 23 SRF cryomodules operating at 2 K, 5 K, and 40 K temperature levels supplied by a single helium cryoplant providing 2.5 kW of cooling capacity at 2.0 K. The PIP-II cryogenic system consists of two major systems: a helium cryogenic plant and a cryogenic distribution system. The cryogenic plant includes a refrigerator cold box, a warm compressor system, and helium storage, recovery, and purification systems. The cryogenic distribution system includes a distribution box, intermediate transfer line, and a tunnel transfer line consisting of modular bayonet cans which supply and return cryogens to the cryomodules. A turnaround can is located at the end of the Linac to turnaround cryogenic flows. This paper describes the layout, design, and current status of the PIP-II cryogenic system.

## 1. Introduction

The Proton Improvement Plan II (PIP-II) is currently being designed and constructed at Fermilab's accelerator complex. It features a new superconducting radio-frequency linear accelerator to provide a powerful high-intensity proton beam (800 MeV, 1.2 MW) for the most intense high energy beam of neutrinos to the international Deep Underground Neutrino Experiment at the Long Baseline Neutrino Facility (LBNF). The PIP-II linear accelerator (Linac) consists of superconducting cryomodules which include a half-wave resonator (HWR) operating at 162.5 MHz, two types of single spoke resonators operating at 325 MHz (SSR1 and SSR2) and two types of elliptical 5-cell cavities at 650 MHz (LB650 and HB650) for a total of twenty-three (23) SRF cryomodules.

Cryogenics plays a vital role in the PIP-II design. The cryogenic system provides cooling to the SRF cavities in the Linac. The cavities operate at 2 K with 5 K and 40 K for cooling of thermal intercepts and the high temperature shield, respectively as well as liquefaction capacity to fill a 15,000 L Dewar to aid in fast cool downs and maintenance activities. The cryogenic plant is being purchased from Air Liquide Advanced Technologies as an in-kind contribution from the Department of Atomic Energy (DAE) India. This in-kind contribution is the single largest contribution within the PIP-II project. The distribution box and tunnel transfer line of the cryogenic distribution system is also a planned in-kind contribution from Wroclaw University of Science and Technology (WUST) as a contribution from Poland. All systems have completed their final design stages and are currently in various procurement and fabrication stages.

## 2. PIP-II cryogenic system

Figure 1 shows a block diagram of the PIP-II cryogenic system which consists of two major systems: the helium cryogenic plant and the cryogenic distribution system (CDS) which collectively provide the required cryogenic cooling capacity to the Linac.

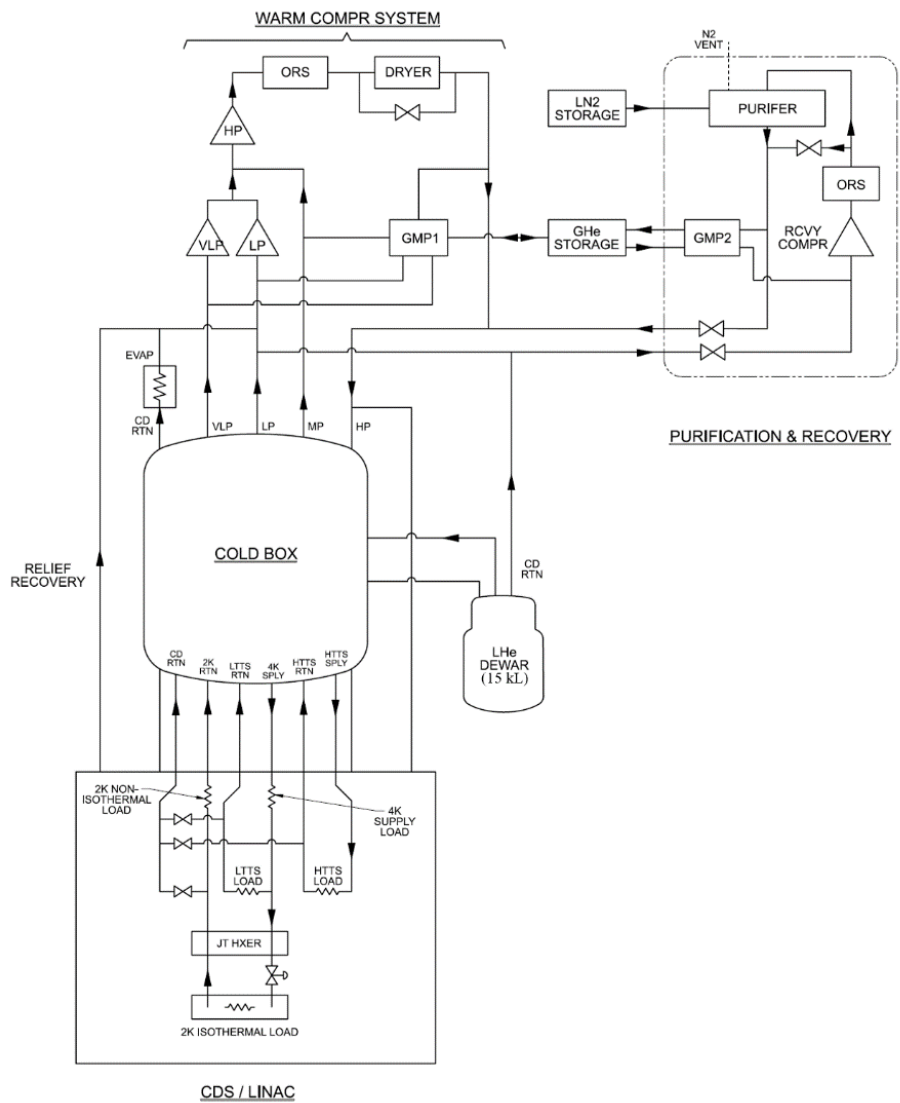

**Figure 1.** Cryogenic system block diagram

The cryoplant includes the warm compression system, the cold box, gaseous and liquid helium inventory, purification, and recovery systems. The cryoplant provides the required cooling capacity at three nominal temperature levels: 40 K for the high temperature thermal shields and intercepts, designated as HTTS; 4.5 K for the low temperature intercepts, designated as LTTS; and 2 K for the SRF cavities and magnets within each cryomodule. The combined 4.5 K supply is divided into two streams as it enters the cryomodules, one that is directed to the Joule-Thompson (JT) heat exchanger preceding the cavity supply and another directed to the LTTS. Each cryomodule also has a dedicated cooldown valve and JT valve to supply cryogenics to each cryomodule string. Table 1 shows the cooling capacity of the supplied cryogenic plant and Figure 2 shows the overall layout of the helium cryogenic plant within the PIP-II complex.

The cryogenic distribution system interfaces with the cold box at the cryoplant building to provide the cryogenic helium supply and return to the cryomodules in the Linac. The CDS consists of one (1) distribution valve box, ~265 m of cryogenic transfer line designated the intermediate transfer line connecting the distribution valve box to the Linac tunnel, twenty-five (25) modular bayonet cans, one for each cryomodule plus two spares at the end for future upgrades, and one (1) turnaround can. The CDS scope also includes the supply and return warm gaseous helium headers for the cooldown and relief systems, as well as the gaseous nitrogen and instrument air headers to the Linac.

Table 1. Cryogenic plant capacity.

| Temperature Level (K) | Capacity (W) |
|---|---|
| 2K Isothermal | $\geq 2500$ |
| 2K Non-isothermal | $\geq 460$ |
| 4.5K Supply | $\geq 180$ |
| LTTS | $\geq 1500$ |
| HTTS | $\geq 10,680$ |

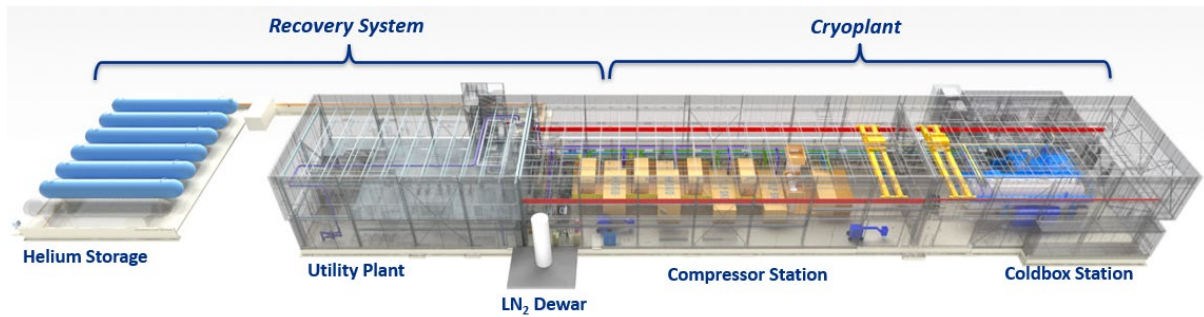

Figure 2. Cryogenic plant layout

## 3. Cold box

The cryogenic system cold box features an integrated design including eight (8) heat exchangers, four (4) turbine expanders and three (3) cold compressors to provide the required cooling capacity without the need for liquid nitrogen precooling. The cold box operates based on four (4) pressure levels: very low pressure (VLP) level, low pressure (LP) level, medium pressure (MP) level and high pressure (HP) level. The helium gas returning from the cold box is first compressed from the VLP and LP pressure level to the MP pressure level through oil flooded warm compressors and combines before being compressed to the HP level by the second stage of compressors.

The HP helium from the warm compressor system enters the cold box at 310 K and ~19 bara which then flows through the heat exchangers, where it is cooled with the MP stream (6 bara), the LP stream (1.1 bara) and the VLP stream (0.4 bara). The HP stream flows through the air adsorber at the 80 K level to remove any air contamination. At 20 K, it is fed to the H2/Ne adsorber to remove any trace amount of hydrogen and neon.

The HP helium stream is partially expanded through a total of four (4) turbine expanders to provide the required cooling for the HP stream at approximately 65 K, 20 K, 11 K and finally the 6 K level. Downstream of the final expander, the HP helium either enters the 15,000 L Dewar when in liquefaction mode or subcooled in a 4.5 K LHe bath to exit as supercritical helium for the cryomodules and the LTTS.

The LTTS helium gas returns to the cold box at approximately 9 K and 2.4 bara. The VLP helium from the 2 K cryomodule load returns to the cold box at above 0.027 bara and approximately 4 K. It is then compressed through a set of three (3) stages of cold compression and through the heat exchangers to recover enthalpy before returning to the VLP compressor. Figure 3 shows the process flow diagram of the cold box.

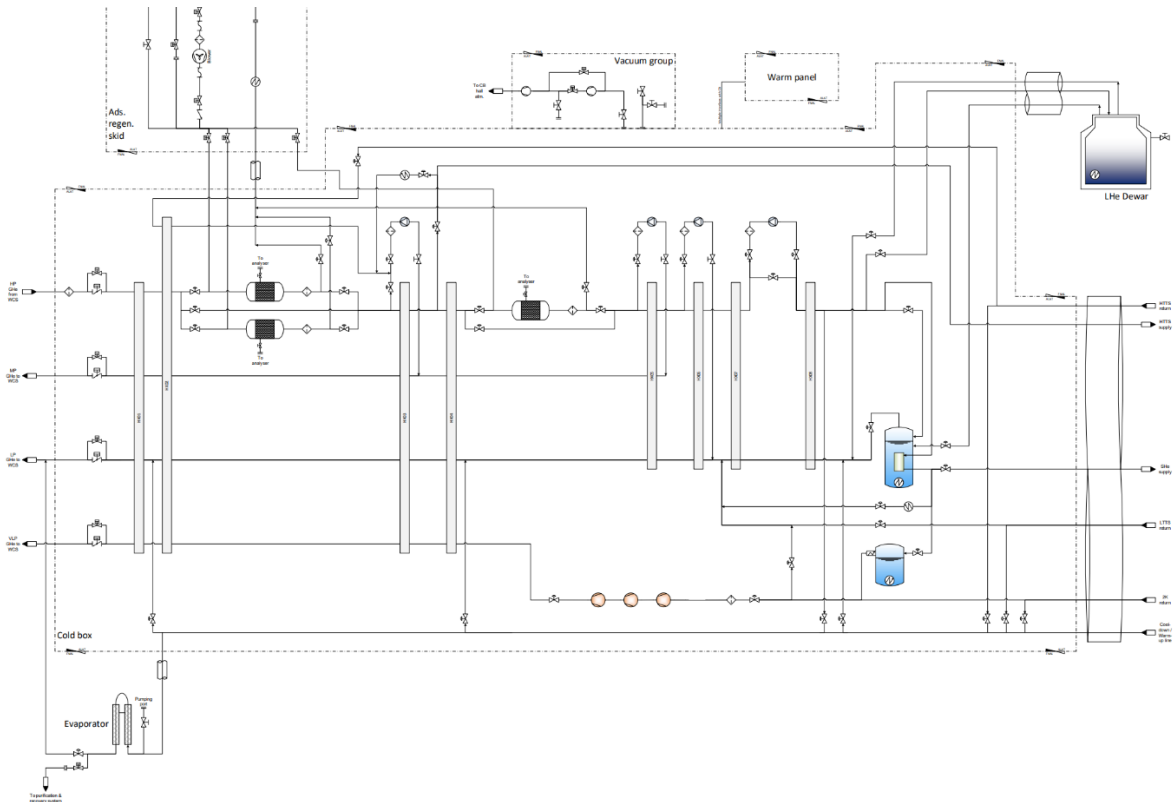

**Figure 3.** Process flow diagram of the cold box

The cold box can support five (5) operational modes: maximum capacity mode, nominal design mode, 4.5 K standby mode, liquefaction mode, and fast cooldown mode. In liquefaction mode, the cold box can produce up to 14.5 g/s of liquid helium. Fast cooldown mode can provide up to 80 g/s to cooldown individual cryomodules from 40 K to 5 K for cryomodule cavity flux expulsion. To allow for a liquid helium buffer volume for cryoplant/cryomodule maintenance and a means for testing performance of each of the major subsystems, a 15,000 L liquid helium Dewar is used. Liquid helium storage is sized to allow for a rapid refill of the cryomodules during fast cooldown and to support maintenance activities.

## 4. Warm compressor system

The main warm helium compressor system consists of three compressor stages manufactured by Howden. The VLP compressor consists of a 561 kW VLP compressor providing 132 g/s from 0.5 bara to 3.05 bara in maximum capacity mode, a 561 kW LP compressor providing 282 g/s from 1.05 bara to 3.05 bara in maximum capacity mode, and two 1384 kW HP compressors providing 894 g/s from 3.05 bara to 19.04 bara in maximum capacity mode.

The HP1 and HP2 compressor skid is composed of two separate compressor bodies and electric motor trains mounted on a single skid. Both compressors share a single oil separator, oil cooler, aftercooler and oil pumping system. The oil pumping system consists of two oil pumps where only one is needed for running of the skid. There are also additional connections for an additional single-skid HP compressor in the future. The LP and VLP compressor skids have their own bulk oil removal system which include oil separator, oil cooler, aftercooler, and oil pump. The LP and VLP compressors share the same compressor body and main drive motor but are mounted and different skids. The valves and piping design between these skids will allow VLP and LP compressors to backup each other during 4.5 K and/or liquefaction mode, where only one compressor is required.

Gas management valves around each of the compressor stages and to the gaseous storage vessels provide system pressure regulation. The oil removal system consists of three (3) coalescers and a charcoal adsorber. The $1^{st}$ and $2^{nd}$ coalescers collect oil that is returned to compressor suction while the $3^{rd}$ coalescer is used as a redundant safeguard. A dryer skid is installed downstream of charcoal adsorber for removing moisture at start up and after maintenance.

## 5. Recovery system

The recovery system includes all the components needed to store, circulate, and purify the helium gas for the entire cryogenic system. The system is also used to recover helium utilizing six (6) helium gas storage vessels, repurposed from the decommissioned Tevatron cryogenic system, with enough volume to store all the helium used in the cryogenic system and cryomodules. Each storage tank is rated at 1.7 MPa with a volume of 114 m$^3$. The gas storage tank farm is connected using multiple gas distribution headers which allow individual storage tank connection to the gas purification system, tube trailer fill station, and the compressor system and the flexibility to purify individual tanks while others are connected to the cryogenic plant. Provisions are also in place to add additional storage tanks in the future with the necessary space and connections.

The purification system consists of a 60 g/s helium gas purifier matched with a single 60 g/s 300 kW Mycom oil flooded screw compressor, also repurposed from the Tevaton cryogenic system, with provisions to add a second compressor as a future upgrade. The compressor includes an oil removal system consisting of three-stages of oil coalescing, a charcoal adsorber, a molecular sieve vessel and a final filter. The purification system is used to purify the entire cryogenic system including the cryomodules using up to 60 g/s of process flow at 1.9 MPa with a purity of less than 1 ppm$_v$ of air and moisture. A 34,000 L liquid nitrogen dewar is used to supply liquid nitrogen to the purifier as well as for cryoplant absorber regeneration and as a source for nitrogen gas for the Linac. The recovery system also includes a gas management system to maintain system pressures and redirect flow to different areas of the system.

## 6. Cryogenic distribution system

The cryogenic distribution system (CDS) consists of the components needed to feed and return the cryogens, via vacuum insulated pipelines to the Linac components needing these services, throughout the entire PIP-II Linac. The CDS includes a distribution valve box, cryogenic intermediate transfer line, tunnel transfer line, and associated helium relief and return headers. A simplified layout of the cryogenic distribution system and Linac is shown in Figure 4. The distribution valve box and tunnel transfer line are planned as in-kind contributions from WUST.

### 6.1. Distribution valve box

The distribution valves box is located in the cryogenic plant cold box room and connects the cold box to the transfer line that feeds the Linac cryomodules. The distribution valve box includes process isolation valves to isolate the cold box from the Linac transfer line. Secondary valves within the cold box allow for double isolation of each circuit. Gas valves allow for purification and warm-up independent of the cold box. The distribution valve box also includes relief valves to account for all CDS relieving scenarios for each cryogenic circuit as well as critical instrumentation such as flow meters for diagnostic purposes. Figure 5 shows the reference design of the distribution valve box. The distribution valves box was designed as a partnership between Fermilab and WUST with the fabrication planned as an in-kind contribution provided by WUST.

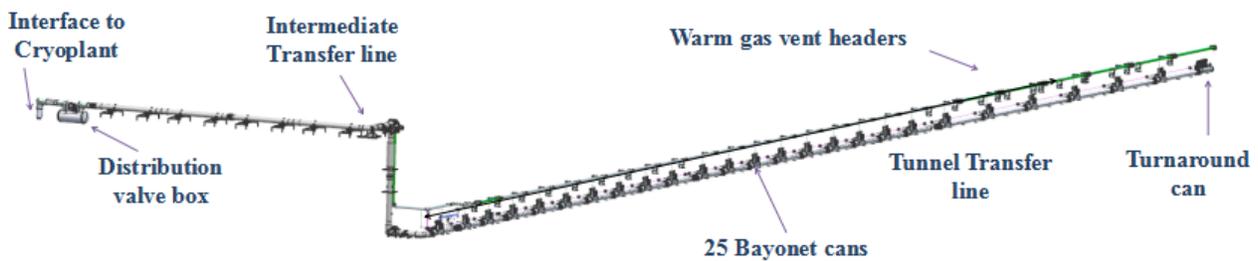

**Figure 4.** Cryogenic distribution system layout

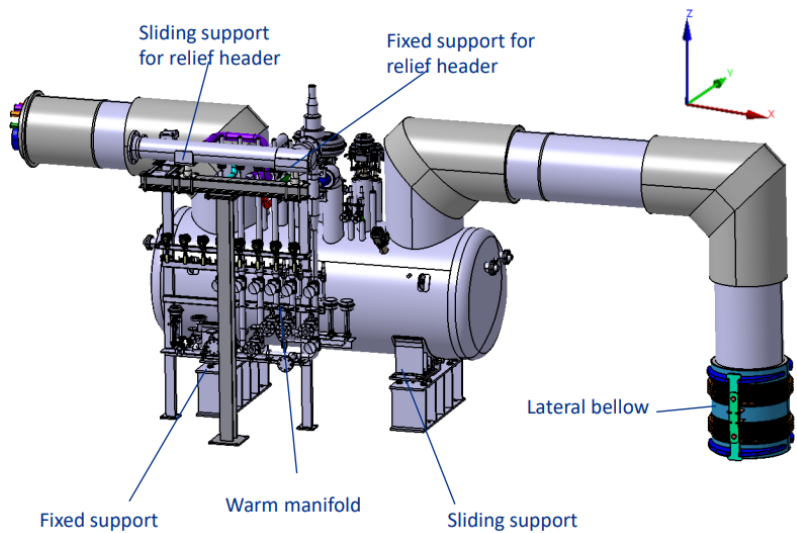

**Figure 5.** Distribution valve box

*6.2. Intermediate transfer line*

The intermediate transfer line (ITL) connects the distribution valve box to the Linac tunnel transfer line. The ITL begins at the distribution valve box in the cold box room extending outdoors along the roof of the technical bay which connects the cryoplant building to the Linac highbay area. The transfer line then enters a vertical chase penetration into the Linac tunnel enclosure. The ITL is currently being procured in industry.

*6.3. Tunnel transfer line*

The tunnel transfer line (TTL) consists of individual connected transfer line modules, one for each individual cryomodule. Each module is a standardized design with different lengths depending on the different cryomodule lengths. In total there are five (5) different lengths; one for each type of cryomodule (HWR, SSR1, SSR2, LB650, and HB650). The tunnel transfer line exchanges cryogenic helium with the cryomodule loads through removable bayonet U-tubes. Each transfer line module includes five female bayonets, one for each circuit (supply and return), eight cryogenic control valves, and associated instrumentation. The bayonet connections allow for positive isolation of individual cryomodules for maintenance while maintaining the cryogenic system at cryogenic temperatures. In addition to the cold process lines, the CDS also provides a warm helium supply for controlled cool down and warm up of the Linac as well as pressure relief vent lines for the cryomodules and the CDS itself. The TTL is designed to allow for fast cool down of the cryomodules. Figure 6 shows the preliminary design of one TTL segment for a cryomodule.

*6.4. Relief system*

CDS relieving for all process circuits is available at each end of the CDS, with a set of reliefs at the distribution valve box and the turnaround can. Cryomodule positive pressure relieving is available at each TTL module. Cryomodule sub-atmospheric 2 K relief devices are directly mounted on each cryomodule. For the 2 K circuit of the cryomodules, a dual pressure rating is used, utilizing the higher allowable stress of the niobium cavities when cold < 80 K.

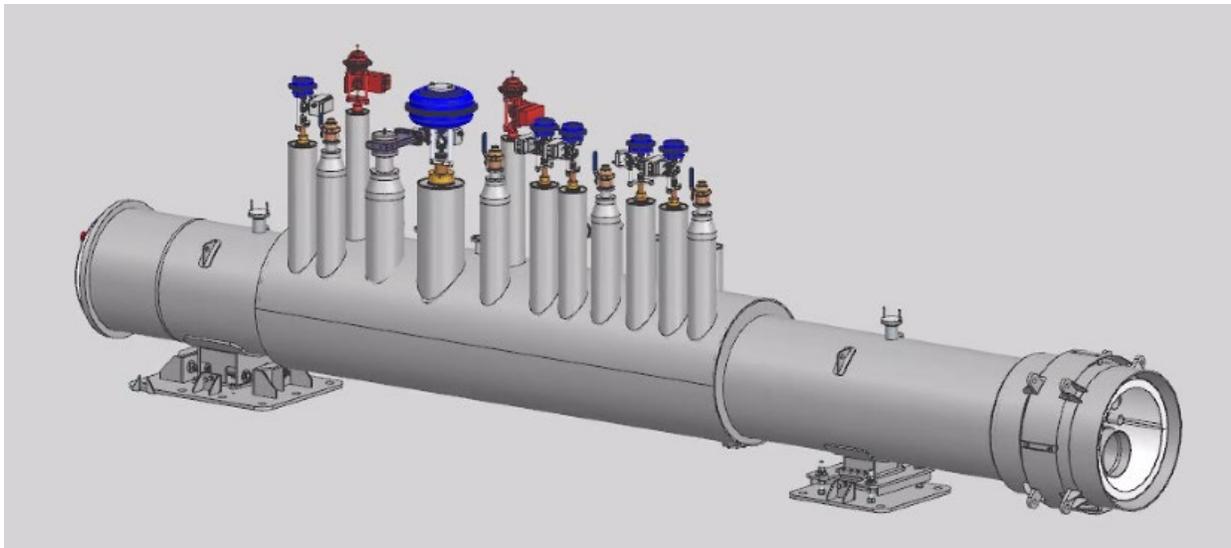

**Figure 6.** Tunnel transfer line module

## 7. Conclusion

The PIP-II cryogenic system is currently under construction at Fermilab which will supply cryogens to the PIP-I Linac, enabling the world's most intense beam of neutrinos to LBNF/DUNE. The cryogenic plant is being provided as an in-kind contribution from DAE India and managed by BARC. This in-kind contribution is the single largest in-kind contribution to the PIP-II project. The distribution valve box and tunnel transfer line reference designs were provided in collaboration with WUST with the fabrication planned also as an in-kind contribution. The PIP-II cryogenic system final designs have been completed and with the various components in the fabrication or procurement stage. The cryoplant building has been constructed and infrastructure installation has begun.

## 8. References

[1]   Jia Y, A, Hansen B, Creus Prats J, Atassi O, Klebaner A Chakravarty A, Goyal M and Kumar J 2022 Status of the PIP-II cryoplant. *IOP Conf. Series: Materials Science and Engineering* **1240** 012066

**Acknowledgments**

This manuscript has been authored by Fermi Research Alliance, LLC under Contract No. DE-AC02-07CH11359 with the U.S. Department of Energy, Office of Science, Office of High Energy Physics. The authors would also like to acknowledge the contributions from our BARC and WUST colleagues as well as the efforts of the APS-TD Cryogenic Technology Division personnel involved in the design and construction of the cryogenic system.